\documentclass[aps,preprint,prd,floatfix,nofootinbib]{revtex4-1}
\pdfoutput=1
\usepackage{amsmath} 
\usepackage{graphicx}
\newcommand{\be}{\begin{equation}} 
\newcommand{\ee}{\end{equation}}
\newcommand{\bea}{\begin{eqnarray}} 
\newcommand{\eea}{\end{eqnarray}}

\begin{document}
\title{Scalar Hair Around Charged Black Holes in Einstein-Gauss-Bonnet Gravity}
\author{Nicol\'{a}s Grandi}
\email{grandi@fisica.unlp.edu.ar}
\affiliation{Instituto de F\'{i}sica de La Plata - CONICET  \&  Departamento de F\'{\i}sica - UNLP,\\ C.C. 67, 1900 La Plata, Argentina}
\author{Ignacio Salazar Landea}
\email{peznacho@gmail.com}
\affiliation{Centro At\'omico Bariloche, 8400-S.C. de Bariloche,
R\'{\i}o Negro, Argentina}
\begin{abstract}
We explore charged black hole solutions in Einstein-Gauss-Bonnet gravity in five dimensions, with a charged scalar hair. We interpret such hairy black holes as the final state of the superradiant instability previously reported for this system. We explore the relation of the hairy black hole solutions with the non-backreacting quasibound states and scalar clouds, as well as with the boson star solutions.
\end{abstract}
\maketitle
\tableofcontents
\newpage
\section{Introduction}
\label{sec:Introduction}
Believed to play a central role in many astrophysical processes, black holes are also interesting as mathematical objects. 
Of particular importance are the uniqueness theorems, stating that the only regular (on and outside a horizon) black hole solution in vacuum is the Kerr-Newman one \cite{Chrusciel:2012jk}. This result holds in 4D and for Einstein-Maxwell dynamics, but it is sometimes generalized as a ``no hair conjecture'' \cite{Ruffini:1971bza}. I asserts that black holes are fundamental objects that should be described only by the externally observable (conserved) parameters: mass, angular momentum, and any additional conserved charge. The physical idea behind the no-hair conjecture is based on an intuitive (and sometimes oversimplified) picture: all matter fields which at the exterior of a black hole would eventually be swallowed by the black hole or be radiated away to infinity, with the exception of fields which are associated with conserved charges.

A series of no hair theorems has been developed for different matter contents in Einstein gravity, including a scalar field \cite{chase}, massive vectors, and spin two fields \cite{Bekenstein:1971hc}. Also, a series of counter examples have been found where some of the assumptions needed for the no hair theorems are violated \cite{Herdeiro:2015waa}. For instance, for asymptotically AdS space-times the AdS works as a box allowing hairy black hole to exists. These were widely studied in the context of holographic superconductors \cite{Hartnoll:2008kx}. 

Within flat space hairy black holes, an open possibility is to study non-linear matter fields where the resistance of the matter field against collapse into the black hole is anchored to these non-linearities \cite{Volkov:1989fi}. Such is the case of black holes with Yang-Mills hair. Another posibility is to study solutions for which the matter field does not inherits the spacetime symmetries. This is the case of Kerr black holes with scalar hair \cite{Herdeiro:2014goa,Herdeiro:2015gia}, where the metric remains stationary and axially symmetric, in the same sense as Kerr, but the matter field  is time dependent.

In five dimensions, Myers-Perry black holes are the natural extension of Kerr black holes, and solutions with scalar hair were contructed in \cite{Brihaye:2014nba}\footnote{Interestingly, considering Skyrme instead of scalar hair one might find spherically symmetric solutions for hairy black holes in five dimensions \cite{Brihaye:2017wqa} }. An interesting feature about five dimensional gravity is that Gauss-Bonnet term becomes dynamically non-trivial. The effect of the Gauss-Bonnet deformation to Myers-Perry hairy black holes was studied in \cite{Brihaye:2015qtu}.

In this paper we study solutions with a complex scalar atmosphere around an Einstein-Gauss-Bonnet charged black hole \cite{Wheeler:1985nh,Banados:2003cz}. In our solutions the metric remain static but the scalar field do not, since it has a time dependent phase. In this sense, our solutions are similar to those in \cite{Herdeiro:2014goa}, but we do not need the black hole to rotate in order to support the scalar hair\footnote{A previous example of spherically symmetric hairy black holes in Einstein-Gauss-Bonnet gravity was studied in \cite{Sotiriou:2014pfa}, where a non-minimal coupling of a real scalar to the Gauss-Bonnet terms is needed in order to obtain regular solutions.}.

The paper is organized as follows. In section \ref{sec:EinsteinGaussBonnetTheory} we present the model whose solutions will be studied. In section \ref{sec:RegularSolutions} we construct its regular spherically symmetric solutions 
 focusing our atention on the boson stars  \ref{sec:BosonStarSolutions}. In section \ref{sec:BlackHoleSolutions} we contruct the black hole solutions to the model, from the charged hairless black hole \ref{sec:TheChargedBlackHole} to the spherically symmetric scalar perturbations including quasibound states and scalar clouds  \ref{sec:ScalarPerturbations}, and finally the hairy black hole solutions \ref{sec:TheHairyBlackHole}. In section  \ref{sec:Discussion} we sutdy the regions of existence of the different solutions in parameter space and a function of the total mass and charge. Finally in section \ref{sec:ConclusionsAndFutureDirections} we explain our conclusions and future research lines. 

\section{Einstein-Gauss-Bonnet theory}
\label{sec:EinsteinGaussBonnetTheory}
The action for Einstein-Gauss-Bonnet theory in the presence of a Maxwell field and
a charged scalar field, reads  
\begin{equation}
I\left[  g_{\mu\nu},A_{\mu},\Phi\right]  =I_{grav}\left[  g_{\mu\nu}\right]
+I_{mat}\left[  A_{\mu},\Phi,g_{\mu\nu}\right]  \ ,
\label{eq:ActionFull}
\end{equation}
where the gravitational part is given by
\begin{equation}
I_{grav}\left[  g_{\mu\nu}\right]  =\frac{1}{2}\int d^{5}x\sqrt
{-g}\left[  R+\alpha  \left(  R^{2}-4R_{\mu\nu}R^{\mu\nu}+R_{\alpha\beta
\gamma\delta}R^{\alpha\beta\gamma\delta}\right)  \right]  \ ,
\label{eq:ActionGravitational}
\end{equation}
while the matter reads
\begin{equation}
I_{matt}\left[  A_{\mu},\Phi,g_{\mu\nu}\right]  =\int d^{5}x\sqrt{-g}\left(
-\frac{1}{4} F_{\mu\nu}F^{\mu\nu}-|D_{\mu}\Phi|^{2}-m^{2}|\Phi|^{2}\right)  \ ,
\label{eq:ActionMatter}
\end{equation}
with $D_{\mu}=\nabla_{\mu}-iqA_{\mu}$. Notice that we have re-absorbed the gravitational coupling into the electric charge of the scalar $q$ by re-scaling the electromagnetic and scalar fields. Here the Gauss-Bonnet coupling $\alpha$ has mass dimension $-2$. Together with the mass of the scalar $m$, these are the only relevant couplings of the theory.

~ 

We look for spherically symmetric stationary solutions of the above defined theory with the general form given by the Ansatz
\bea
&&
ds^{2}=-N^2f   dt^{2}+\frac{dr^2}{f}+r^{2}d\Omega_{3}^{2}\,,
\label{eq:MetricAnsatz}
\\
&&A=h\, dt \,, 
\label{eq:GaugeFieldAnsatz}
\\
&&\Phi=e^{-i \omega t} \phi  \,,  
\label{eq:ScalarAnsatz}
\eea
where $N, f, h$ and $\phi$ are functions of $r$, and $\omega$ is the frequency of the scalar field. Notice that the time dependence of the scalar field is a phase, which implies that the resulting energy momentum tensor and electric current are time-independent. This ensures the compatibility of the Ansatz. The resulting equations of motion read
\bea
&& 
N'=
\frac
{2 r^3 \left(f^2 N^2 \phi'^2+\phi^2 (\omega +q h)^2\right)}
{3 f^2 N \left(4 \alpha(1\!-\!f)+r^2\right)}\,,
\label{eq:lapse}\\
&&
f'=-\frac{fN'}{N}+\frac
{ 
	r 
	\left(
		2 N^2 
		\left(
			3(1\!-\!f)-r^2 m^2\phi^2
		\right)
		-
		r^2 h'^2
	\right)
	}
{3  N^2 \left(4 \alpha (1\!-\!f)+r^2\right)}\,,
\label{eq:radial}\\
&&
fN\left(\frac{r^3 h'}{N}\right)'
=2   q r^3  (\omega+qh)\phi^2\,,
\label{eq:maxwell}\\
&&
f N
\left(r^3 f N \phi' 
\right)'
=
r^3 \left(m^2 f N^2-(\omega +q h)^2\right) \phi\,.
\label{eq:scalar}
\eea
In asymptotically flat space, these equations imply at large $r$ the asymptotic forms
\bea
N&\approx&1+\dots\, ,
\label{eq:AsymptoticLapse}\\
f&\approx&1- \frac{M}{r^2} + \dots\,, 
\label{eq:AsymptoticRadial}\\
h&\approx&\mu- \frac{Q}{r^2} +\dots  \,,
\label{eq:AsymptoticMaxwell}\\
\phi&\approx&  \phi_\infty \frac{ e^{\pm\sqrt{m^2-\omega^2}r}}{r^{3/2}}+  \dots\,.
\label{eq:AsymptoticScalar}
\eea
Here $Q$ and $M$ are the charge and mass of the solution, and $\mu$ is an arbitrary constant of integration, that we set to zero in what follows. 

~ 

{Equations \eqref{eq:lapse}-\eqref{eq:scalar} have few analytic solutions. To solve them numerically, we use a shooting method implemented in Mathematica. We expand the functions close to the origin/horizon in powers ot $r$, to identify the independent parameters. Using {\tt NDSolve}, we integrate the equations towards a large distance cutoff to obtain the asymptotic behavior. Finally, we adjust the parameters to mach the desired decays \eqref{eq:AsymptoticLapse}-\eqref{eq:AsymptoticScalar} by using {\tt FindRoot}.}

\section{Regular solutions}
\label{sec:RegularSolutions}
Regular solutions  to the theory (\ref{eq:ActionFull}) with asymptotic conditions \eqref{eq:AsymptoticLapse}-\eqref{eq:AsymptoticScalar} are defined as solutions in which the metric and scalar field are regular everywhere. In particular, they are regular close to the origin $r=0$. By expanding the scalar field equation at small $r$, we get
\bea
&&
\left(r^3  \phi' 
\right)'
\approx
r^3 \left(m^2 -\left(\frac{\omega \!-\! \omega_0}{N_0}\right)^2\right) \phi\,,
\label{eq:NearOriginScalar}
\eea
where we defined the constants $N_0=N(0)$ and  $\omega_0=-qh(0)$. Notice that, in order to avoid a conical singularity at the origin, we have set $f_0=f(0)=1$. By analyzing the different solutions to this equation, we can study the different regular configurations of the system

\subsection{Flat solution}
\label{sec:FlatSolutions}
Equation \eqref{eq:NearOriginScalar} has the trivial solution $\phi=0$ {\em i.e.} the scalar field and its first radial derivative vanish close to the origin. This is a trivial solution for the complete scalar equation \eqref{eq:scalar}, implying that \eqref{eq:lapse} is solved by $N=N_0$, and \eqref{eq:maxwell} by $qh=-\omega_0$, which in turn implies that $f=1$ in eq.  \eqref{eq:radial}. This is the trivial empty flat space solution.

\subsection{Scalar perturbations around flat space}
\label{sec:ScalarPerturbationsStar}
Since the right hand side of equations \eqref{eq:lapse}-\eqref{eq:radial} is quadratic in the scalar field $\phi$, the flat space metric $N=f=1$ still a good solution of the system even for non-vanishing scalar field, as long as $|\phi|$ is small enough. In such ``probe'' approximation, we call the resulting solutions ``scalar perturbations''. 

Scalar perturbations satisfy the equation for the scalar field \eqref{eq:scalar} particularized for the flat space background, that takes the form \eqref{eq:NearOriginScalar}. This is simply the scalar wave equation in spherical coordinates. Close to the origin, it is solved by
\bea
\phi\approx\phi_0+\frac{ 1}8\left(m^2 -\left(\frac{\omega \!-\! \omega_0}{N_0}\right)^2\right) \phi_0 r^2+\dots\,.
\label{eq:BoundarConditonScalarStar}
\eea
This behavior implies the absence of any sink or source for the scalar field. Then propagating $\omega^2>m^2$ modes cannot be sustained. On the other hand, bound states $\omega^2<m^2$ would constitute solutions of the kind known as ``Q-balls''. As shown in \cite{Coleman:1985ki}, in order for them to exist an additional attractive potential must be added to the Lagrangian. This is very reasonable from a physical point of view: neither free $q=0$ nor repelling $q\neq0$ particles could condense in the absence of any attractive interaction.

\subsection{Boson star solutions}
\label{sec:BosonStarSolutions}
A natural question emerges from the above analysis: could the role of the attractive interaction needed for the scalar condensation to take place be played by gravity?
Boson stars are defined as a regular spherically symmetric self gravitating solutions to the theory (\ref{eq:ActionFull}) with asymptotic conditions \eqref{eq:AsymptoticLapse}-\eqref{eq:AsymptoticScalar} \cite{Liebling:2012fv}.  In Gauss-Bonnet theory they were first constructed in \cite{Hartmann:2013tca} for the $q=0$ case. Here we present solutions with $q\neq0$.

To obtain the boundary conditions at the origin for the metric and gauge fields, we start by plugging the expansion \eqref{eq:BoundarConditonScalarStar} for the scalar field into \eqref{eq:maxwell}, to get 
\bea
&&
 qh 
\approx
-\omega_0+\frac{q^2\phi_0^2}{4}   (\omega-\omega_0) r^2+\dots\,.
\label{eq:BoundaryConditionMaxwellStar} 
\eea
With this, consistency of the remaining equations \eqref{eq:lapse} and \eqref{eq:radial} implies the expansion for the lapse and radial functions
\bea
f&\approx&1+\frac{r^2}{4 \alpha}\left(1
-
\sqrt{1+\alpha\frac{4\phi_0^2}{3N_0^2} \left( \left( \omega-\omega_0\right)^2+m^2 N_0^2      \right)   } \right)+\dots\,,
\label{eq:BoundarConditionRadialStar}
\\
N&\approx&N_0+ \frac{ \left( \omega-\omega_0\right)^2\phi_0^2}
{3N_0\sqrt{1+\alpha \frac {4\phi_0^2}{3N_0^2}\left(\left( \omega-\omega_0\right)^2
+m^2N_0^2  \right)}} r^2+\dots\, .
\label{eq:BoundaryConditionLapseStar}
\eea

~ 

\begin{figure}[t]
\begin{center}
\includegraphics[width=3.5in]{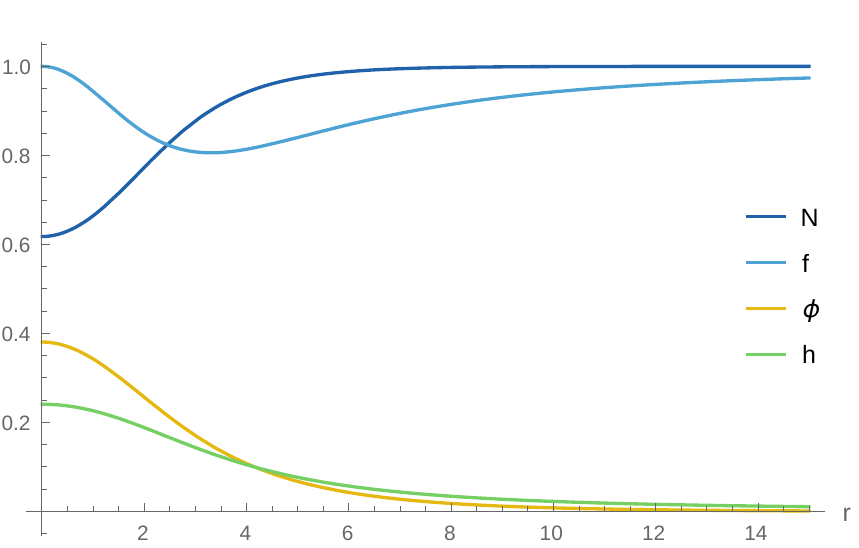}
\caption{\label{fig:BosonStarProfile}
Typical profiles for the lapse $N$, the radial function $f$, the electric potential $h$, and the scalar $\phi$, for a boson star solution with $q=\alpha=0.5$. 
}
\end{center}
\end{figure}

We integrated numerically the equations of motion \eqref{eq:lapse}-\eqref{eq:scalar} with boundary conditions given by \eqref{eq:BoundarConditonScalarStar}-\eqref{eq:BoundaryConditionLapseStar}. A typical profile for the boson star is shown in Figure \ref{fig:BosonStarProfile}.

\begin{figure}[t]
\flushleft\includegraphics[width=3.in]{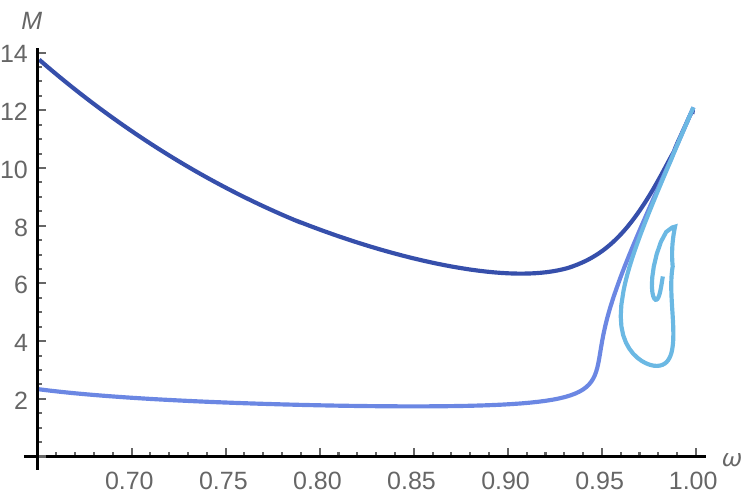}  \quad \includegraphics[width=3.in]{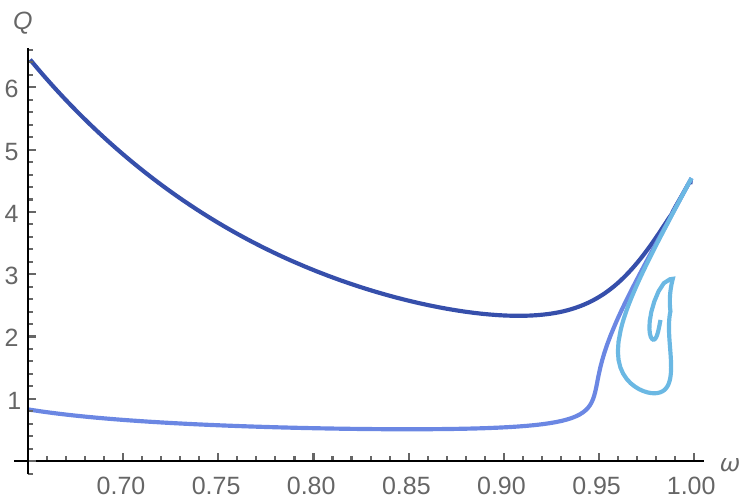}
\\
\center{\qquad\qquad\qquad\qquad\qquad\includegraphics[width=3.in]{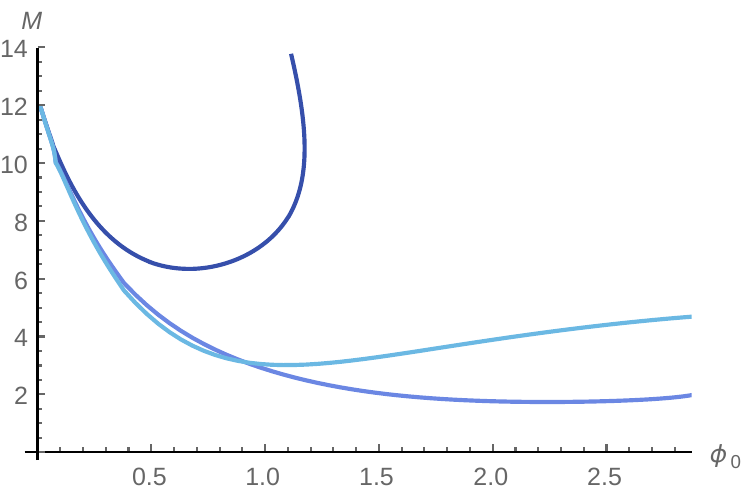} \qquad \qquad \includegraphics[width=0.8in]{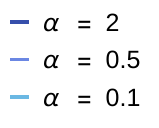}}
\caption{
\label{fig:BosonStarExistence} Top: ADM mass (Left) and charge (Right) as functions of the value of the scalar frequency $\omega$ for $q=0.5$ and different values of $\alpha$.
Bottom: ADM mass as a function of the value $\phi_0$ of the scalar field at the origin for $q=0.5$ and different values of $\alpha$.
}
\end{figure}

In order to explore the space of solutions, we vary the value of the scalar at the origin. At the bottom of Figure \ref{fig:BosonStarExistence} we show how the mass of the star depends on $\phi_0$ at fixed $q$. We observe that we can have arbitrarily large central densities when the Gauss-Bonnet parameter $\alpha$ is small, while there is a bound in the central density when $\alpha$ is large enough.

In order to find regular star solutions we need, for a given central density, to fix a value for the frequency. At the top of Figure \ref{fig:BosonStarExistence} we plot the star mass $M$ and charge $Q$ as functions of the frequency of the scalar field at fixed $q$ for different values of $\alpha$. This soutions are the same apearing at the bottom of Figure \ref{fig:BosonStarExistence}, but now we focus on $\omega$ instead of $\phi_0$. Solutions exist for $\omega^2<m^2$ as expected from the asymptotic behavior. As we decrease the frequency, we find two different behaviors depending on the value of $\alpha$. For small values of $\alpha$ we see that the behavior is quite similar to that of Einstein boson stars. Solutions exists down to certain minimal value of $\omega$ and then bounce up and down in $\omega$ in a spiraling pattern. As we increase the Gauss-Bonnet parameter, we see that the spirals unwrap and the spiraling behavior dissapears. A similar phenomenology was reported for the ungauged case in \cite{Hartmann:2013tca}.
\section{Black hole solutions}
\label{sec:BlackHoleSolutions}
Black hole solutions are obtained by assuming that the function $f(r)$ vanishes at a certain radius $r_h$ that defines the black hole horizon. By expanding the scalar equation of motion around such radius, we get the approximated form
\bea
(4\pi T)^2(r\!-\!r_h)
\left((r\!-\!r_h)\phi' 
\right)'
\approx
\left( 
4\pi T N_hm^2(r\!-\!r_h)-(\omega\!-\!\omega_s)^2\right)\phi  \,,
\label{eq:NearHorizonScalar}
\eea
where we defined the black hole temperature as $4\pi T=f'(r_h)N(r_h)$, the superradiant frequency as $\omega_s=-qh(r_h)$, and the constant $N_h=N(r_h)$. This equation controls the near horizon behavior of the scalar field. By analyzing it, we can explore the different black hole solutions of the theory.
\subsection{The charged black hole}
\label{sec:TheChargedBlackHole}
As a first example, we notice that eq. \eqref{eq:NearHorizonScalar}
admits solutions with a vanishing $\phi$. This implies that, close to the horizon, both $\phi$ and its radial derivative $\phi'$ vanish. Since the complete scalar equation \eqref{eq:scalar} is a second order linear equation, such trivial boundary condition at the horizon implies that the scalar vanishes everywhere. The equation \eqref{eq:lapse} for $N$ is trivial and, together with the boundary condition at infinity \eqref{eq:AsymptoticLapse}, implies that $N=1$ everywhere. The remaining equations \eqref{eq:radial} and \eqref{eq:maxwell} can be solved analytically to obtain the charged black hole solution, as
\bea
&&N=1\,,
\label{eq:BHLapse}
\\
&&f(r)=1 +\frac{r^2}{4 \alpha} \left(1-\sqrt{1+\alpha \left(
\frac{8 M}{ r^4}-\frac{16 Q^2} {3 r^{6}}\right)}\right)\,,
\label{eq:BHRadial}
\\
&&h=-\frac{Q}{r^2}\,,
\label{eq:BHMaxwell}
\\
&&\phi=0\,,
\label{eq:BHScalar}
\eea
where $M$ and $Q$ are constants of integration parametrizing to the mass and charge of the black hole respectively. They are related with the horizon radius $r_h$ by the constraint $f(r_h)=0$, implying
\bea
r_h^2=\frac12\left(
{M}-2\alpha+ \sqrt{\left( M-2\alpha \right)^2-\frac{8}{3} Q^2}     
\right)
\,.
\label{eq:HorizonRadius}
\eea
As it is evident in this formula, the existence of a horizon puts a constraint in the allowed values of the charge and mass for any given value of the Gauss-Bonnet parameter $\alpha$.
\subsection{Scalar perturbations of the charged black hole}
\label{sec:ScalarPerturbations}
Again, since the right hand side of equations \eqref{eq:lapse}-\eqref{eq:maxwell} is quadratic in the scalar field amplitude $\phi$, the probe approximation works and the charged black hole solution \eqref{eq:BHLapse}-\eqref{eq:BHMaxwell} still a good solution of the system even for non-vanishing scalar field, as long as $|\phi|$ is small enough. 

The scalar perturbations satisfy the scalar equation \eqref{eq:scalar} particularized to the background \eqref{eq:BHLapse}-\eqref{eq:BHMaxwell}, namely
\bea
f 
\left(r^3 f \phi' 
\right)'
\approx
r^3 \left(m^2 f-(\omega +q h)^2\right) \phi\,.
\label{eq:ScalarPerturbations}
\eea

In order to fix the boundary conditions for the scalar perturbations, we go back to the formula \eqref{eq:NearHorizonScalar}, that in the present case has $N_h=1$ (or equivalently we expand eq. \eqref{eq:ScalarPerturbations} close to the horizon). We see that the behavior  as $r$ approaches $r_h$ depends on whether $\omega=\omega_s$ or $\omega\neq\omega_s$. This allows us to classify the scalar perturbations according to their frequency, as the ``quasibound states'' with $\omega\neq\omega_s$, and the ``scalar cloud'' with $\omega=\omega_s$.
\subsubsection{The quasibound states and the superradiant instability}
\label{sec:SuperradiantInstability}
For the modes with $\omega\neq\omega_s$, for $r-r_h$ small enough we can drop the term linear in $r-r_h$ in the right hand side of \eqref{eq:NearHorizonScalar}, to obtain
\bea
(4\pi T)^2(r-r_h)
\left((r-r_h)\phi' 
\right)'
\approx
-(\omega-\omega_s)^2 \phi\,.
\label{eq:NearHorizonScalarQuasinormal}
\eea
This allows for the simple oscillating solution
\bea
\phi\approx \phi_\pm(r\!-\!r_h)^{\pm i\frac{\omega-\omega_s}{4\pi T}}+\dots\,.
\label{eq:BoundaryConditionQuasinormal}
\eea
In order to obtain a physically meaningful solution, we need to have an expression that is well behaved at the horizon. Since information carried by wave packages cannot go out of the black hole, the group velocity $v_g$ has to be in-going. Since $v_g=\pm4\pi T$ this implies that we must chose the solution with the ``$-$'' sign.

The so-defined quasibound states have a phase velocity given by $v_f=4\pi T Re(\omega)/(\omega_s\!-\!Re(\omega))$. This is negative for $Re(\omega)>\omega_s$, implying that the energy transported by such modes is flowing into the black hole, and then they are being damped, {\em i.e.} they have $Im(\omega)<0$. On the other hand, it becomes positive for modes with $Re(\omega)<\omega_s$, implying that such modes are extracting energy out of the black hole, and then they are being amplified, having $Im(\omega)>0$. This entails an instability whenever $\omega_s\neq0$, known as the ``superradiant'' instability. 

{ In a slightly different context, the superradiant instability of this system was studied in \cite{Fierro:2017fky}. There, the scalar field modes
were confined inside a perfect mirror of finite radius at which the scalar field vanishes. The resulting discretized frequencies have a imaginary part that moves from negative to positive values when the radius of the mirror is made large enough. }

\begin{figure}[t]
\begin{center} 
\includegraphics[width=3.2in]{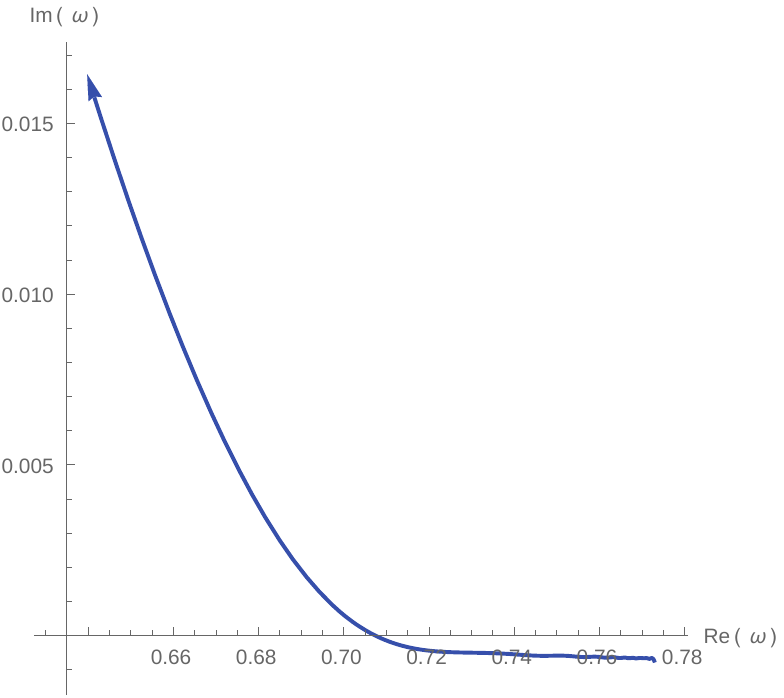}\hfill \includegraphics[width=3.2in]{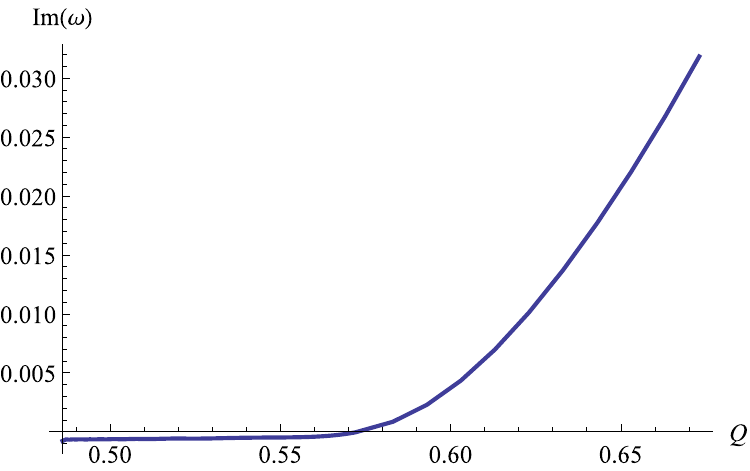}
\caption{\label{fig:qnfreq} 
Left: The frequency of the fundamental quasibound state on the complex plane, for $M=5.08$, $\alpha=2$ and varying the black hole charge $Q$. The arrow points in the direction of growing charge, we see that for $Q$ large enough the imaginary part becomes positive, signaling an instability. Right: Imaginary part of the quasinobound state a for black holes with $M=5.08$, as a function of the black hole charge $Q$. Again we choose $\alpha=2$. At $Q\approx 0.575$, $Im(\omega)$ changes sign, trigering an instability. 
}
\end{center}
\end{figure}

In our present context of quasibound states that decay at infinity, in order to get regular solutions we must again quantize the frequency. The resulting fundamental frequency is plotted in Figure \ref{fig:qnfreq} for a fixed black hole mass $M$ as a function of the black hole charge $Q$. As we increase the black hole charge we go from the stable regime $Im(\omega)<0$ into the superradiant instability $Im(\omega)>0$. The unstable modes increase exponentially as time goes by, eventually making the probe approximation no longer valid. A natural expectation is that the end point of this instability would be a hairy configuration. Indeed, interpolating from positive to negative  $Im(\omega)$ implies going through stable solutions with real frequencies. These are of particular interest since they have zero energy flux across the horizon. We study them the forthcoming section.

\subsubsection{The scalar cloud}
\label{sec:TheScalarCloud}
In view of the analysis above, one could expect that there must be a particular quasibound state that sets the frontier between the decaying stable modes and the amplified superradiant ones.  Indeed, for the particular mode with $Re(\omega)=\omega_s$ and $Im(\omega)=0$, the term linear in $r-r_h$ in the right hand side of \eqref{eq:NearHorizonScalar} dominates, and the equation reads
\bea
(4\pi T)(r\!- \!r_h)
\left(( r \!-\!r_h)\phi' 
\right)'
\approx
 m^2 (r\!-\!r_h)\phi\,.
\label{eq:NearHorizonScalarCloud}
\eea
This is solved by a power series expansion of the form
\bea
\phi\approx\phi_h +\frac{m^2}{4\pi T} \phi_h (r\!-\!r_h) +\dots\,.
\label{eq:BoundaryConditionScalarCloud}
\eea
In terms of an undetermined constant $\phi_h$. This entails a different boundary condition not included in \eqref{eq:BoundaryConditionQuasinormal}, giving rise to a different kind of solutions. 

\begin{figure}[h]
\begin{center}
\includegraphics[width=3.2in]{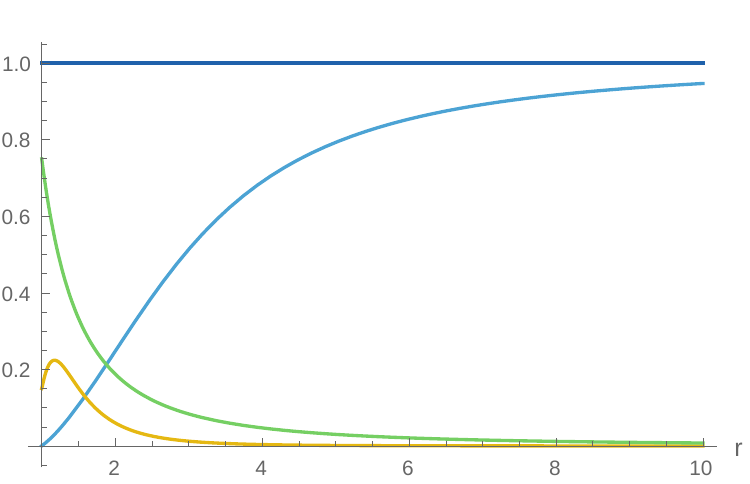}\includegraphics[width=3.2in]{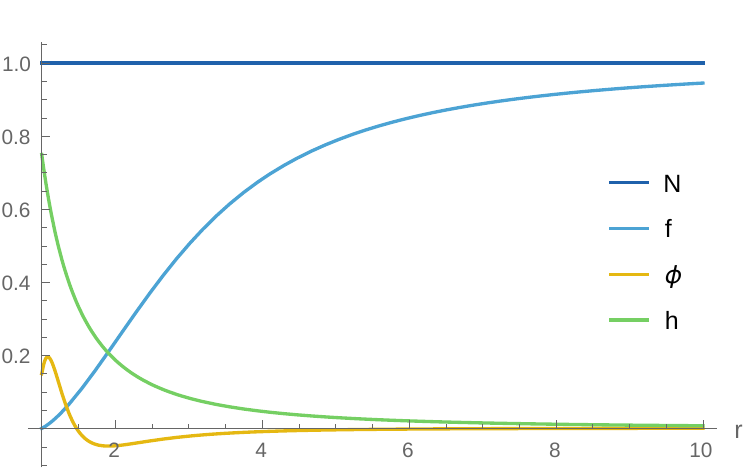}
\caption{
\label{fig:ScalarCloudProfile0} 
Top: Typical profile of a scalar cloud without nodes.
It corresponds to a scalar cloud sitting around a black hole with mass $M=5.376$ and charge $Q=0.751$. Bottom: Typical profile of a scalar cloud with one node. It corresponds to a scalar cloud sitting around a black hole with mass $M=5.520$ and charge $Q=0.884$. The profiles for the functions $f$, $N$ and $h$ are those corresponding to the charged black hole solution.
}
\end{center}
\end{figure} 

We can integrate the complete scalar perturbations equation \eqref{eq:ScalarPerturbations} with boundary conditions given by \eqref{eq:BoundaryConditionScalarCloud} to obtain the profile of the so called ``scalar cloud'' \cite{Hod:2012px}.
Since the frequency is now fixed, it cannot be adjusted to quantized values in order to satisfy regularity conditions at infinity. This implies that the scalar cloud exists only for a restricted subset of parameter space. We argue that such scalar cloud is the end point of the superradiant instability, as long as the probe approximation remains valid.

On the left panel of Figure \ref{fig:ScalarCloudProfile0} we show a typical profile of a scalar cloud, corresponding to a solution without nodes for the scalar field. Notice that the validity of the probe approximation is limited by the maximum value of the scalar. On the right panel of Figure \ref{fig:ScalarCloudProfile0} a scalar cloud with one node is shown.

In our numerical calculations we checked that, as the frequency $\omega_s$ approaches the mass $m$, the black hole charge needed to support the scalar cloud diverges. This could have been expected from the asymptotic behavior.
\subsection{The hairy black hole}
\label{sec:TheHairyBlackHole}
\begin{figure}[t]
\begin{center}
\includegraphics[width=3.5in]{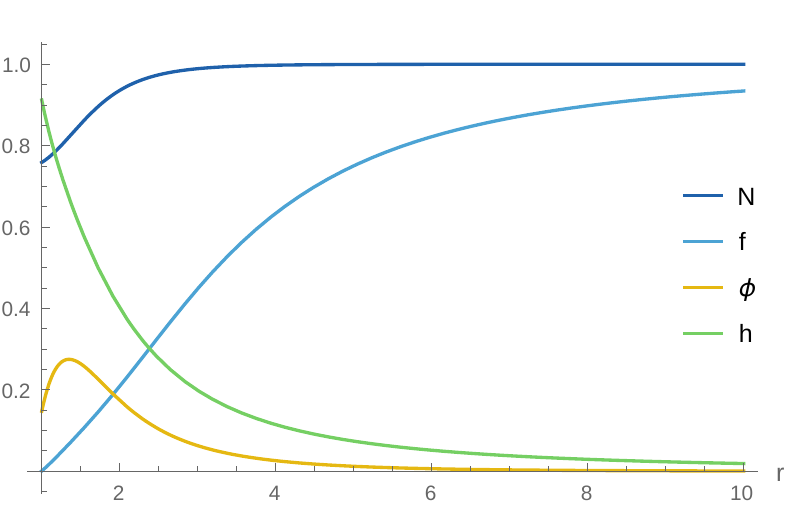}
\caption{\label{fig:HairyBlackHoleProfile} Typical profiles for the lapse $N$, the radial function $f$, the electric potential $h$ and the scalar $\phi$, for a hairy black hole solution.}
\end{center}
\end{figure}

The cases studied in the previous subsection suggest the following question: if the scalar cloud can be considered as the end point of a superradiant instability in the probe limit, then what would be its backreacting form?  In order to answer it, we go back to the equation for the behavior of the scalar field in the near horizon region \eqref{eq:NearHorizonScalar} and put $Re(\omega)=\omega_s$ and $Im(\omega)=0$, obtaining
\bea
(4\pi T)(r\!-\!r_h)
\left((r\!-\!r_h)\phi' 
\right)'
\approx
 N_hm^2(r\!-\!r_h)\phi  \,.
\label{eq:NearHorizonScalarFully}
\eea
This is solved by
\bea
\phi\approx\phi_h +\frac{m^2N_h}{4\pi T} \phi_h (r\!-\!r_h) +\dots\,,
\label{eq:BoundaryConditionScalarHairy}
\eea
in terms of the undetermined constants $\phi_h, N_h$ and $T$. This assumes the following expansion of the radial and gauge functions
\bea
f &\approx & \frac{4\pi T}{N_h}(r\!- \!r_h)+\dots\,,
\label{eq:BoundaryConditionsRadialHairy}
\\
qh &\approx & -\omega + qh'_h(r\!-\!r_h)+\dots\,,
\label{eq:BoundaryConditionsMaxwellHairy}
\eea
with a new constant $h'_h$. Notice that we have imposed $-qh(r_h)\equiv\omega_s=\omega$. The consistency of the equation of motion \eqref{eq:lapse} at the horizon requires 
\bea
N \approx N_h+ \frac{2r_h^3\phi_h^2(m^4N_h^2+q^2{h'_h}^{\!2})}{3(4\pi T)^2N_h^3(4\alpha+r_h^2)}(r\!-\!r_h)+\dots\,,
\label{eq:BoundaryConditionsLapseHairy}
\eea
where in order to satisfy \eqref{eq:radial} the black hole temperature is defined as
\bea
4\pi T = r_h\frac{2N_h^2(3-r_h^2m^2\phi_h^2)-r_h^2{h'_h}^{\!2}}{3N_h^2(4\alpha+r_h^2)}\,.
\label{eq:TemperatureHairy}
\eea 
We have three free parameters in the near horizon expansion of the fields $h'_h,\,N_h,\,\phi_h$. These will be used as shooting variables when integrating numerically the equations to find solutions with the desired asymptotics. A typical profile for the functions $f,N,h$ and $\phi$ is shown in Figure \ref{fig:HairyBlackHoleProfile}. 

\section{Discussion}
\label{sec:Discussion}
In this section we discuss the regions of existence for the different solutions presented in the previous sections. 

\begin{figure}[t]
\begin{center} 
\includegraphics[width=4.5in]{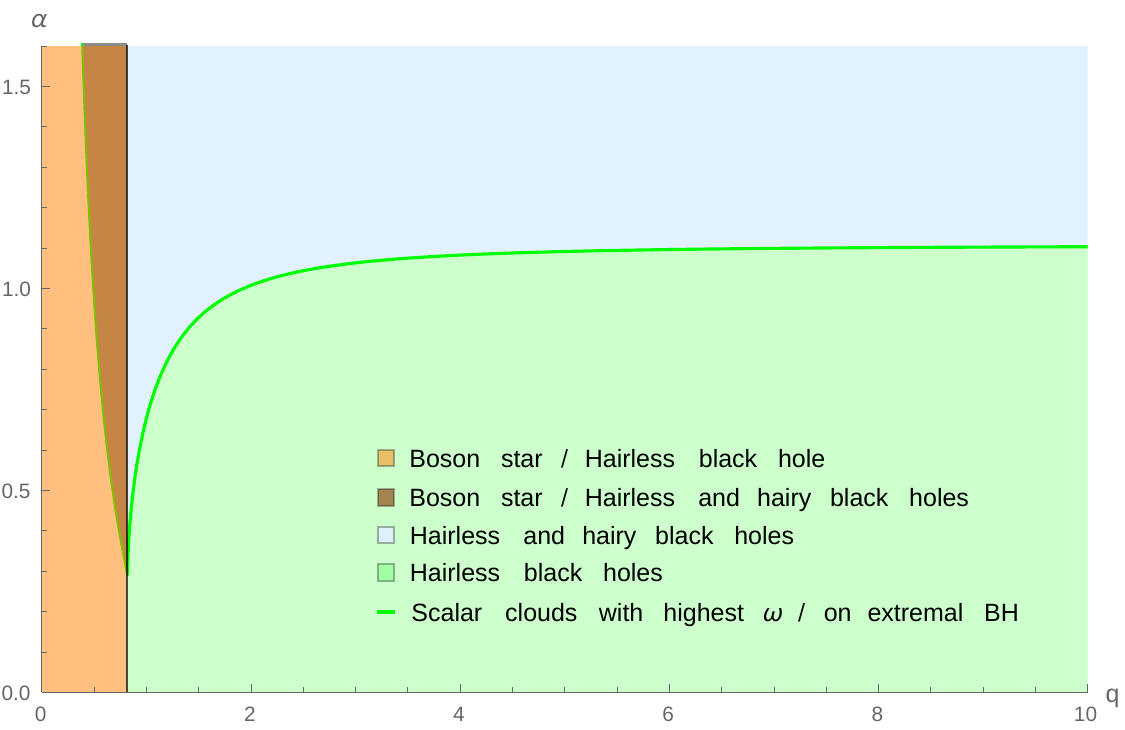}
\caption{\label{fig:RegionsOfExistence} 
Regions with and without hairy black hole soltions in the $q~ vs.~\alpha$ plane. We see that the existence of the hairy solution requires a nonvanishing value of the Gauss-Bonnet parameter $\alpha$ and a lower bound on the electromagnetic coupling $q$. On the other hand, the existence of the boson star solution is possible only if $q$ lies bellow a different bound. Charged black holes exists on the whole plane. For the sake of numerics we set $m^2=1$ and $r_h=1$.  }
\end{center}
\end{figure}

We start by analyzing the space of possible theories. It is spanned by the parameters $q$ and $\alpha$.  A plot of such parameter plane is shown in Figure \ref{fig:RegionsOfExistence}. At any point in the plane, the trivial empty flat space solution exists. The same is true for the charged black hole solution.

On the flat space background, spherical scalar perturbations do not exist, neither in the form of a spherical wave nor as a Q-ball. The latter would require the existence of some sort of attractive interaction between scalar particles \cite{Coleman:1985ki}. When backreaction is taken into account, gravity plays the role of such attractive force, allowing the scalar field to condense, forming a boson star. However, backreaction also turns on the electromagnetic repulsion between particles, proportional to the parameter $q$. As $q$ grows, the repulsion becomes stronger and, when a a critical value is reached, the condensate dissipates. Naively, one would expect boson star solutions only for ${q^2}\leq {m^2}/2$, {\em i.e.} when the gravitational force between two scalar particles is bigger than the electromagnetic force  \cite{Jetzer:1989av}. Nevertheless, when the non-linearities of the equations of motion are taken into account, one can find solutions even in the ${q^2}> {m^2}/2$ regime \cite{Pugliese:2013gsa}. In particular, our numerics show that solutions exist up to $q^2\approx 0.67m^2$, which correspond to the orange and brown regions shown in Figure \ref{fig:RegionsOfExistence}. The value of $q$ at which the star ceases to exist is independent of $\alpha$ up to our numerical precision. 
 
On the charged black hole background, scalar perturbations can be turned on. They exist in the form of quasinormal modes and quasibound states. Superradiant $Re(\omega)<\omega_s$ quasibound states are known to be present \cite{Fierro:2017fky} in this theory. A particular quasibound state with $\omega=\omega_s$ sets the frontier between superradiant and stable modes, and it represents the scalar cloud. The largest possible frequency for a scalar cloud is $\omega=\omega_s\equiv m$, and these clouds live on a line on the $\alpha~ vs.~ q$, represented by the green line on the right in Figure \ref{fig:RegionsOfExistence}. Regarding the green line on the left, it represents a scalar cloud sitting on an extremal black hole. In the intermediate light blue and brown regions we can always find scalar clouds for some values of the mass and charge of the black hole.  These two green lines meet at a particular point $q^2=\frac23 m^2$.

It is interesting to note that, up to numerical precision, there is a remarkable coincidence of the boundary of the region in which boson stars exist $q^2\simeq 0.67m^2$, and the cusp of the line at which scalar clouds exist $q^2=\frac{2}{3}m^2$ at small $\alpha$. This can be understood as follows: on the outer region of a boson star, the scalar field forms an ``atmosphere''  that is in equilibrium with the gravitational and electromagnetic fields sourced by a central configuration with mass $M$ and charge $Q$. Such fields are indistinguishable of those sourced by a hairless black hole with the same mass $M$ and charge $Q$.
In consequence, if the charge $q$ is such that no black hole can sustain a scalar cloud, then the boson star would be unable to sustain its scalar atmosphere, becoming unstable. Since the metric of the boson star is not highly curved, its boundary of existence must be almost independent of the Gauss-Bonnet parameter, and coincide with the smallest $\alpha$ scalar cloud.   In this sense the bound $q^2<\frac23 m^2$ can be considered as the correction of the bound presented in  \cite{Jetzer:1989av} when considering the general relativistic and classical field theory effects.

Now, if we consider backreaction, we expect that the scalar cloud would turn into a hairy black hole. This is indeed what happens: the hairy black hole solution exists in the light blue and brown regions of Figure \ref{fig:RegionsOfExistence}. 
We see that hairy black hole solutions require a finite value of $\alpha$, in other words they do not exist in General Relativity without the Gauss-Bonnet deformation. Moreover, the hairy solutions exists only for a finite value of the electric charge. We can interpret this minimal value as the electromagnetic repulsion needed to compensate the gravitational collapse.  Notice the coincidence of the boundary of the light blue region with one line of existence of scalar clouds. This suggests that the hairy black hole is the end point of the superradiant instability.  Since both boson stars and hairy black hole solutions exist in the brown region of Figure \ref{fig:RegionsOfExistence}, one may conjecture that in that region hairy black holes might be the end point of a boson star collapse.



\begin{figure}[t]
\begin{center}
\includegraphics[width=4.5in]{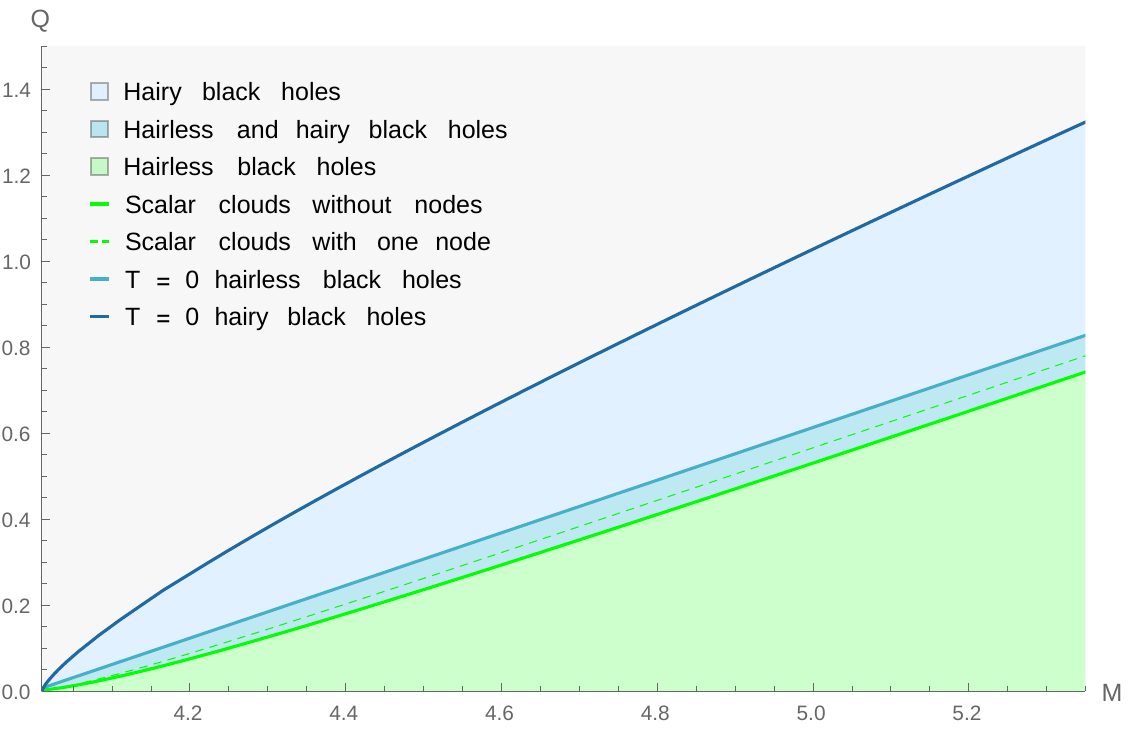}\hfill
\caption{\label{fig:RegionsOfExistence2} Region of existence of the different solutions in the $Q~vs.~M$ plane, for $q=m=1$, $\alpha=2$. Notice that the hairy black holes solutions begin to exist exactly at the curve in which there are scalar cloud solutions without nodes.}
\end{center}
\end{figure}

Let us now fix the theory at particular values of $q$ and $\alpha$ at any point inside the region of existence of hairy black hole solutions. With this, we can explore the domain of existence of our different solutions with respect to the value of the integration constants $Q$ and $M$. In Figure \ref{fig:RegionsOfExistence2} we show such $Q~vs.~ M$ plane for $\alpha$ and $q$ fixed. 

We have no solutions in the gray region. The blue line represents the extremal $T=0$ hairy black hole. Bellow such line, the light blue region  contains only hairy black holes. At the light blue curve, extremal $T=0$ charged black holes without hair exist. Bellow it in the darker blue region both kinds of black holes are present. In this region we have a degeneration of solutions bypassing the no hair conjecture. Moreover, in its interior the hairless black hole solution can be dressed with a scalar cloud with a certain number of nodes. For example, the dotted green line represents the existence of a scalar cloud with one node. Interestingly, the continuous green curve, representing the frontier for the existence of hairy black holes, coincides with the curve at which nodeless scalar clouds can be found. In view of that, we claim that the hairy black hole is the end point of the superrandiant instability, since the line of existence for hairy solutions coincide with the line of appearence of superradiant modes.

\section{Conclusions and future directions}
\label{sec:ConclusionsAndFutureDirections}
%
%
Along this paper we studied spherically symmetric solutions to the Einstein-Gauss-Bonnet-Maxwell-Klein-Gordon theory. Considering different boundary conditions we found several different solutions. A first classification comes from wether the geometry has an events horizon or not
\begin{itemize}
\item Horizonless regular solutions with a non-trivial profile for the scalar field correspond to boson stars. 
\item Solutions with an event horizon correspond to black hole, and can be classified according to the behavior of the scalar field at the horizon. 
\begin{itemize}
\item A trivial profile for the scalar field gives a charged black hole, whose analytic form is well known. 
\item Considering a probe scalar arround the black hole background with in-falling boundary conditions, we get the quasibound states. These have a complex characteristic frequency that imply that there is a flux of energy at the horizon. For some black hole solutions we find that, as we increase the black hole mass $M$, the $Im(\omega)$ changes sign from negative to positive, giving rise to a superradiant instability. 
\item In the same probe approximation, precisely at $Im(\omega)=0$ we find a particular kind of solutions dubbed scalar clouds, where we have no flux of energy accross the horizon. This turn out to be a good approximation to light hairy black holes,
\item Finally, we found hairy black hole solutions, {\em i.e.} solutions where we solved numerically the full non-linear system of equations.
\end{itemize}
\end{itemize}
These last fully non-linear solutions are the main result of this paper. We studied their region of existence in parameter space and compared them with the other aformentioned solutions.
%
%
As future directions, one possibility would be to study the linear stability of the hairy black hole solutions. Furthermore, we have for simplicity considered the minimal dimension for which the Gauss-Bonnet term is non-trivial, it would be interesting to generalize these solutions to arbitrary space-time dimension.

In order to understand how robust is the presented mechanism to generate black hole hair, it would be helpful to study charged Gauss-Bonnet black holes in the presence of a scalar field with a more general potential. Furthermore, on may consider extend our results to spin-one hair, and consider solutions of charged Proca clouds \cite{Sampaio:2014swa}, hair \cite{Herdeiro:2016tmi} and stars \cite{Brito:2015pxa,Garcia:2016ldc}.

Finally, it would be interesting to construct rotating solutions, and see how the solutions presented here connect to those studied in \cite{Brihaye:2015qtu}. 
\section*{Acknowledgements}
\label{sec:acknowledgements}
This work is partially supported by grants PIP-2008-0396 (Conicet, Argentina) and PID-2013-X648 and PID-2017-X791 (UNLP, Argentina). We thank Fede Garc\'ia and Prof. Julio Oliva for helpful exchange.

\end{document}